\documentclass[a4paper,12pt]{article}
\usepackage{graphicx}
 \usepackage{amsmath}
 \usepackage{amsfonts}
 \usepackage{mathtext}
 
\begin{document}

\newfont{\cyrF}{wncyr10 scaled 1440} 
\newcommand{\cyr}{\baselineskip12.5pt\cyrfnt\cyracc}

\title{
On relation between  bulk, surface and curvature parts of nuclear binding  energy  within the model of hexagonal clusters}

\author{V.  V. Sagun$^{1, 2}$, K.  A. Bugaev$^{2, 3}$ and A. I. Ivanytskyi$^{2, 4}$}

\maketitle
{\small
\noindent
$^{1}$ \quad CFisUC, Department of Physics, University of Coimbra, 3004-516 Coimbra, Portugal
\\
$^{2}$ \quad Bogolyubov Institute for Theoretical Physics of the National Academy of Sciences of Ukraine, 03680 Kyiv,  Ukraine
\\
$^{3}$ \quad Department of Physics, Taras Shevchenko National University of Kyiv, 03022 Kyiv, Ukraine
		\\
$^{4}$ \quad Department of Fundamental Physics, University of Salamanca, 
		37008 Plaza de la Merced s/n, Spain		
}

\abstract{Using the model of hexagonal clusters we express the surface, curvature and Gauss curvature
coefficients of the nuclear binding energy in terms  of its bulk coefficient. Using the derived values
of these coefficients and a single fitting parameter we are able to reasonably well describe the experimental binding energies of nuclei with more than 100 nucleons.  To improve the description of lighter nuclei 
we introduce the  same  correction for all the coefficients.  In this way we determine the 
apparent values  of  the surface, curvature and Gauss curvature
coefficients which  may be used for infinite nuclear matter equation of state. This simple model allows us to fix  the
temperature dependence of all these coefficients, if the temperature dependence  for the bulk term is known. The found estimates for 
critical temperature are well consistent   both with experimental and with theoretical findings. }

\section{Introduction}\label{secintro}

For several decades the statistical multifragmentation model (SMM) \cite{SMM} was a good guide for theoretical and experimental studies of the nuclear liquid-gas phase transition. However, at present some of its major assumptions do not look well justified and have to be reexamined
in view of new theoretical developments. Among them we would like to mention that inclusion of the Fisher exponent $\tau$ \cite{Fisher} into 
a simplified version of SMM \cite{SMM1} led to an understanding of  a possible complicated structure of the nuclear liquid-gas phase transition on the basis of exact analytical solution \cite{SMM2,SMM3} of such a simplified model. Examination of the critical exponents of a simplified SMM  made in  \cite{SMM3} led to a conclusion  that, in contrast to the Fisher droplet model  \cite{Fisher},  
the nuclear liquid-gas phase transition 
may  have not a critical endpoint, but the tricritical endpoint, if  the value of exponent $\tau$ is between 1 and 2, while for $\tau \ge 2$ the critical endpoint maybe absent at all \cite{SMM2,SMM3}. 

The concept of surface tension induced by the interaction of  nuclear fragment with the thermal medium suggested in \cite{IST14}
and developed further in \cite{IST18b,IST17a}  allowed us to introduce into the SMM the equation of state of compressible nuclear 
liquid and to successfully account  for the effects of the surrounding medium on the surface tension  of nuclear fragments of arbitrary size in
a way which obeys the L. van Hove axioms of statistical mechanics \cite{Axioms1,Axioms2}.  In addition, this  concept again brings up 
two questions of principal importance for theoretical modeling  of nuclear liquid-gas phase transition: (I) why the curvature contribution of the 
binding energy of large nuclei is absent in the SMM?; and (II) what is the temperature dependence of the proper  surface and curvature parts of the nuclear binding energy  to which the contributions generated by the interaction of  nuclear fragment with the surrounding thermal medium should be added to? 

There is an extended literature in which these main questions are discussed on the basis of different approaches and, hence, we have to apologize that here we quote only a few works  directly related to our discussion. It will mainly concern the relation between the  bulk $-a_V A$, surface $a_S A^\frac{2}{3}$ and curvature $a_C A^\frac{1}{3}$ terms of the binding energy on symmetric  nuclei consisting of  $A\ge 10$ nucleons.  The  empirical Bethe-Weizs$\ddot{\rm a}$cker formula \cite{BW} successfully describes  the binding of large nuclei and the major part of this energy is determined by the  bulk and surface terms which are, respectively, proportional to the coefficients $a_V$ and  $a_S$, while the curvature coefficient $a_C$ in this formula  is set to zero. Over last five decades 
this formula was essentially improved \cite{BW2,BW3,Brack,Dudek, Kolom08,Mor:2012,Kolom11} to account for various properties of  more than 2700 known nuclei.  In some references even two curvature terms, namely the usual one $a_C A^\frac{1}{3}$ and the Gauss one $a_G A^0$, 
are discussed \cite{Brack,Dudek}. In fact,  already in 1953 on the basis of the Fermi gas model Hill and Wheeler concluded  that a curvature dependent term $a_C A^\frac{1}{3}$  should exist in the liquid-drop model of nuclei \cite{Curvat53}, but until now there is no consensus 
on its presence in the  modified Bethe-Weizs$\ddot{\rm a}$cker formula. Moreover, the existing versions of SMM completely ignore it and in part this  can be explained by the hope  that the curvature term may play a minor role in nuclear multifragmentation studies. 
However,  in the realistic extensions of Fisher droplet model  \cite{NewFisher1,NewFisher2}  which are used to describe the liquid-gas phase transition the curvature term is present.  Therefore,  the curvature term should be preset in the SMM of nuclei, but in this case the temperature dependence of surface and curvature terms should be derived simultaneously in order to have a thermodynamically consistent treatment. 
Unfortunately,  within the mean-field approximation  such a task is rather complicated even for the surface tension \cite{Brack,FiniteT}, but  for the statistical model
like the SMM  the results of  mean-field approach cannot be used, since  they  break down the L. van Hove axioms of statistical mechanics 
\cite{Axioms1,Axioms2} and, hence,  will destroy the main attractive feature of SMM, i.e. the  statistical character of phase transition. Therefore, to resolve the two main questions formulated above we need a sufficiently  simple model  which, nevertheless, will not lead to the conflict with the L. van Hove axioms of statistical mechanics. 

A recent paper \cite{Mor:2012} is devoted to an interesting discussion of the relation between the bulk and surface terms in the Bethe-Weizs$\ddot{\rm a}$cker formula. For  this purpose the   authors of Ref. \cite{Mor:2012} considered a toy model of atomic nucleus which have a cubic shape and in which  the  nucleons are also cubic. Accounting for the interaction of  nearest
neighbors one ends up with  the relation $a_V=a_S$ within the cubic model.  At first 
glance this result  cannot be robust due to strongly oversimplified  treatment  of the real physical nuclei within
the model of cubic nuclei. However, even this primitive picture allows us to conclude about existence of 
a strong correlation between the values of  $a_V$ and $a_S$ coefficients in the Bethe-Wiezsacker formula. 

Hence, a partial success  of the cubic nucleus model analyzed  in \cite{Mor:2012}   motivates us to formulate  more
 elaborate model of hexagonal cluster \cite{Mack:1962}.  Using an exact mathematical representation for the number
 of  spherical  particles in each filled layer we will express the coefficients $a_S$, $a_C$ and $a_G$ in terms of a
 single parameter $a_V$. The derived parameterization of the bulk, surface and curvature terms will be compared
 with two most successful fits  of the experimental binding energies of nuclei of  $A \ge 50$ nucleons and with other theoretical  predictions for the coefficient $a_C$. To apply the derived model to the description of  light nuclei with $A\ge 10$ nucleons,
 we will study the corrections of each coefficient in the spirit of leptodermous expansion.  In this  way we will determine the 
 apparent values of the coefficients $a_S$, $a_C$ and $a_G$  which should be used in the modified SMM. Using the fact that
 the same relation between the coefficients $a_V$,  $a_S$, $a_C$ and $a_G$ is valid for finite temperatures, we will obtain 
 the temperature dependence of these coefficients for a single large nucleus in a vacuum and estimate the  critical  temperature 
of such nucleus  above which it becomes absolutely unstable. 

The work is organized as follows. In Section 2 we establish the hexagonal cluster model of nuclei and 
express the coefficients $a_S$, $a_C$ and $a_G$ in terms of the bulk coefficient $a_V$. Section 3 is devoted to 
a discussion of theoretical predictions for the curvature coefficient $a_C$. The same correction
is  introduced for all these coefficients and  their apparent values are found. Also in this Section we 
establish the temperature dependence of the coefficients $a_V$, $a_S$, $a_C$ and $a_G$
using the SMM parameterization of  the bulk coefficient temperature dependence.
Our conclusions are formulated in Section 4. 

\section{Surface and curvature energy of hexagonal clusters}

As was mentioned above, a simple model developed in the paper \cite{Mor:2012} is rather 
unphysical. The main reason of this is  that neither real nuclei nor nucleons are cubic.
In order to study the relation between bulk, surface and curvature terms in the
leptodermous expansion of the Bethe-Weizs$\ddot{\rm a}$cker formula a more realistic model is required.
Here we develop the geometrical model which assume the nuclei to be a hexagonal structures 
of spherical nucleons.  The main object of our analysis will be the main binding energy of nucleus consisting of $A$ nucleons
\begin{eqnarray}
\label{EqIK}
E_B (A, \{a_{K}\}) = - a_V \,A+ a_S\,A^\frac{2}{3} + a_C\,A^\frac{1}{3} \,,
\end{eqnarray}
which describes the bulk $a_V$, surface $a_S$ and curvature $a_C$ terms in the Bethe-Weizs$\ddot{\rm a}$cker formula \cite{BW}.
In particular,  in our analysis we will use the two sets of  Ref. \cite{Mor:2012}
\begin{eqnarray}
\label{EqIIK}
{\rm \bf set\,\,I:} & \Leftrightarrow & E_B^I (A) = E_B (A,  a_V = 15.6 \,{\rm MeV}; ~ a_S= 17.32\,{\rm MeV}; ~ a_C\, = 0\,{\rm MeV}) \,,\\
\label{EqIIIK}
{\rm \bf set\,\,IV:} & \Leftrightarrow &  E_B^{IV} (A) = E_B (A,  a_V = 15.26 \,{\rm MeV}; ~ a_S= 15.26\,{\rm MeV}; ~ a_C\, = 3.6\,{\rm MeV}) \,,
\end{eqnarray}
which provide an excellent fit of the experimental binding energies of nuclei for $A \in [20; 250]$.
Since these sets are in a good  agreement  with the other fits of experimental nuclear binding energies (see, for instance,  \cite{Dudek} for a comprehensive review and the references therein), in what follows  we will accept these 
sets as the two typical representatives which equally well reproduce the experimental data for the nuclei with 
 $A \in [50; 250]$ nucleons. Visually these two curves for $E_B (A)$  can be hardly  distinguished from each other and,
 hence,  we consider them equivalent to each other for the nuclei with 
 $A \in [50; 250]$ nucleons. 

\begin{figure}[ht]
\hspace*{5.5cm}
\includegraphics[height=5 cm,width=5.5cm]{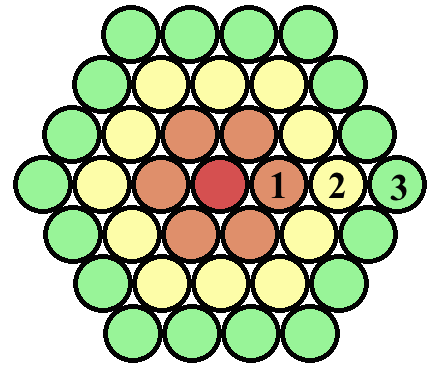} 
\caption{[Color online] The hexagonal structure of cluster  which consists of central particle  (dark circle at the center)
and three covering layers (marked with the digits 1, 2 and 3).}
\label{fig1}
\end{figure}

Similarly to ordinary liquids \cite{Cells} one may think of  the hexagonal clusters consisting  of the cells
in which a single particle (nucleon) is moving.  The cells are formed dynamically by the other nucleons.  
The advantages of this model compared to cubic one are as follows. 
First of all, the shape of the cells  is nearly  spherical. At the same time the large nuclei have a shape 
which is approximately spherical (not cubic!), and this  is in agreement with the common wisdom of modern nuclear 
physics. In addition the hexagonal structure of nuclei in the present model leads to their dense 
 packing which  is also in conformity with the contemporary state of knowledge.  Hence, a hexagonal nucleus can be considered  as a central nucleon cell covered by several  layers of the other cells in which a single nucleon is moving (see Fig. \ref{fig1}).

In what follows we use an approximation of the nearest neighbors interaction. Namely, an interaction of 
 each pair of  nucleons contributes  the energy $\varepsilon<0$ to the binding energy of nucleus. The 
mathematical basis of our  treatment  is the formula which relates the number of
covering layer $k$ ($k=1$ for first layer, $k=2$ for second layer and so on) to the number of
nucleons in it $10k^{2}+2$ \cite{Mack:1962}. For example, the number of nucleons in the 
first layer is 12. In what follows we will consider a large nuclei  with $n \gg1$ covering 
layers. Moreover, similarly to Refs. \cite{Mor:2012,Kolom11} we neglect the effects
 related to the shell structure of clusters
and consider $n(A)$ as a continuous function of number of nucleons $A$.
 The total number of nucleons (mass number) in such a nucleus is
\begin{eqnarray}
\label{EqI}
A=1+\sum^{k=n}_{k=1}\left(10k^{2}+2\right)=
\frac{10n^{3}}{3}+\frac{10n^{2}}{2}+\frac{22n}{6}+1.
\end{eqnarray}
For $A\gg1$ the third and fourth terms in Eq. (\ref{EqI}) are negligibly small comparing to the first and the second ones. Hence we can write
\begin{eqnarray}
\label{EqII}
A\simeq\frac{10n^{3}}{3}\left[1+\frac{3}{2n}\right] \,.
%
\end{eqnarray}
From Eq. (\ref{EqII}) it is convenient  to express the number of outer  (surface)  layer  $n$ as
\begin{eqnarray}
\label{EqIII}
n(A) & \simeq & \left[ \frac{3A}{10}\right]^\frac{1}{3}\left[1+ \frac{3}{2\, n}\right]^{-\frac{1}{3}}
\simeq 
\left[ \frac{3A}{10}\right]^\frac{1}{3}\left[1- \frac{1}{2\, n}\right] \simeq  \nonumber \\
& \simeq &
\left[\frac{3A}{10}\right]^\frac{1}{3}\left[1- \frac{1}{2}\left[\frac{3A}{10}\right]^{-\frac{1}{3}}\right] = \left[\frac{3A}{10}\right]^\frac{1}{3} - \frac{1}{2}
 \,,
\end{eqnarray}
where on the second step of derivation an approximation for the  third power root was applied and 
the leading order approximation $n\simeq(3A/10)^\frac{1}{3}$ was used   on the third step.  A more refined expression for $n(A) \simeq \left[\frac{3A}{10}\right]^\frac{1}{3} - \frac{1}{2} -\frac{7}{60}\left[\frac{3A}{10}\right]^{-\frac{1}{3}} + 0.1005 \left[\frac{3A}{10}\right]^{-\frac{2}{3}}$ provides the maximal absolute  deviation from the physical solution $n(A)$ of Eq. (\ref{EqI}) less than $10^{-3}$ for any $A$ larger than 2.5.

Using  Eq. (\ref{EqIII})
we are able to calculate the binding energy $E_B(A)=-a_V A+a_SA^\frac{2}{3}+a_CA^\frac{1}{3}$ of the
 nucleus under consideration. Each   nucleon not located on the surface  has 12 nearest neighbors.
 At the same time each  nucleon located on the surface ($n$-th layer) of the nucleus  has three vacant positions which
are not filled by nucleons. 
Therefore,  from the bulk energy of the system 
\begin{eqnarray}
\label{EqIV}
- a_{V}A=  12\varepsilon A\,,
\end{eqnarray}
one has to subtract the surface energy of three absent nearest neighbors.
The latter nucleons belong to $(n+1)^{th}$ covering layer and, hence, their 
number is $10\left(n+1\right)^{2}+2$. Thus, one gets
\begin{figure}[t]
\centerline{\includegraphics[height=99mm,width=99mm]{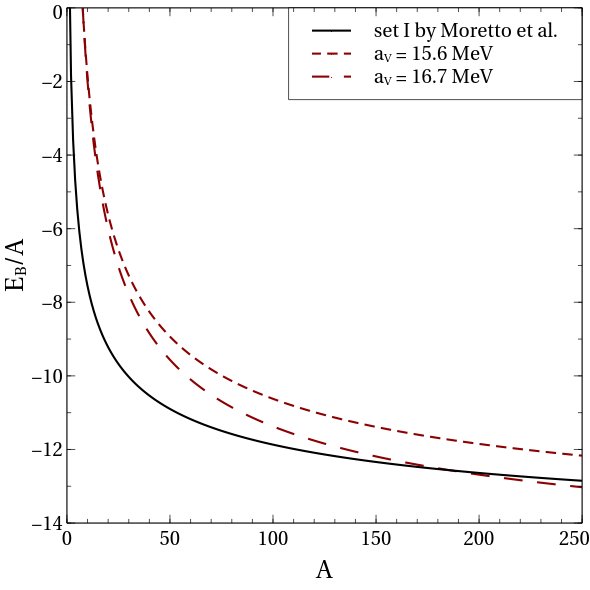} }
\caption{Comparison of the binding energy per nucleon of set I  (solid black curve) with the hexagonal model
for $a_V = 15.6$ MeV (short dashed curve) and  $a_V = 16.7$ MeV (long dashed curve).}
\label{fig2}
\end{figure}
\begin{eqnarray}
\label{EqV}
a_SA^\frac{2}{3}+a_CA^\frac{1}{3}&=&
- 3\varepsilon \left[10\left(n(A) +1\right)^{2}+2\right]= - 3\varepsilon \left[10\left(\left[\frac{3A}{10}\right]^\frac{1}{3} + \frac{1}{2}\right)^{2}+2\right]= \nonumber \\
&=& - 3\varepsilon \left[10 \left[\frac{3A}{10}\right]^\frac{2}{3} + 10 \left[\frac{3A}{10}\right]^\frac{1}{3}+\frac{9}{2}\right]
\,.
\end{eqnarray}
It is remarkable, 
that in contrast  to the model of cubic nuclei of Ref.  \cite{Mor:2012}  the curvature term naturally appears in the present 
model. It is linear in $A^\frac{1}{3}$ whereas the surface term is quadratic. Finally, using the leading order
 approximation  we can find that
\begin{eqnarray}
\label{EqVII}
a_SA^\frac{2}{3}
&\simeq& - 30\varepsilon\left[\frac{3A}{10} \right]^\frac{2}{3}\,, ~ a_CA^\frac{1}{3}
\simeq - 30\varepsilon\left[\frac{3A}{10} \right]^\frac{1}{3}\\
\label{EqVIInew}
E_B^{th} &\simeq& 12\varepsilon A  - 30\varepsilon\left[\frac{3A}{10} \right]^\frac{2}{3} - 30\varepsilon\left[\frac{3A}{10} \right]^\frac{1}{3}.
\end{eqnarray}
The obtained results demonstrate two advantages of the present model. First, in contrast to the cubic  model of Ref. \cite{Mor:2012} one can derive in the same way not only the surface term, but the both curvature terms. Second, in contrast to the models of nuclear forces the present one  does not depend on the definition of nuclear radius $R_A \simeq r_0 A^\frac{1}{3}$ and on the value of particle number density of nuclear ground state. 
In what follows we, however,  neglect the correction $-\frac{27}{2} \varepsilon = - 1.125 a_V$ (the last term in Eq. (\ref{EqV})) which is called the Gauss curvature term \cite{Dudek} in order to present our main results. However, in appropriate places we will comment on how one  can  account for such a correction.

Using Eqs. (\ref{EqIV}) and (\ref{EqVII})  we can  find that in the model
of nuclei with hexagonal structure
\begin{eqnarray}
\label{EqVIII}
\frac{a_{V}}{a_S}&=&\frac{4}{90^\frac{2}{3}}\simeq\frac{1}{1.1204}\,,\\
\frac{a_{S}}{a_C}&=& {0.3^\frac{1}{3}} \simeq 0.6694  \simeq \frac{1}{1.4939}\,.
\label{EqIX}
\end{eqnarray}
Amazingly,  
taking the typical value  of the binding energy per nucleon for zero temperature 
$a_V =16$ MeV used in the statistical multifragmentation model  \cite{SMM},  we obtain that the surface free energy  coefficient is 
$a_S = 1.1204\, a_V =17.9248$ MeV, which is just 0.4 percent less than the typical 
value 18 MeV  used in this profound model \cite{SMM}.  
Taking the bulk term $a_V\biggl|_{sI} =15.6$ MeV of set I of Ref.  \cite{Mor:2012}, from 
Eq. (\ref{EqVIII}) one finds  $a_{S} \simeq 17.477$ MeV which differs from the corresponding value $a_{S}\biggl|_{sI} \simeq 17.32$ MeV  found  in  \cite{Mor:2012}  by about 0.9\% only. 

On the other hand the  value of the curvature term found  in  \cite{Mor:2012} $a_C \simeq \frac {3.6}{15.26} a_S \simeq 0.24 \, a_S$  for the set IV is essentially lower than the result of Eq. (\ref{EqIX}). From  Fig. \ref{fig2} one can see that  the derived curvature term does not provide a good description of the set I 
binding energy. Moreover, the increased  bulk term value $a_V = 16.7$ MeV does not essentially improve the description quality for $A< 50$ 
 as it is seen from  Fig. \ref{fig2}. 
To quantify the mean deviation from the set I  per nuclear mass number  let us introduce the following integral 
\begin{eqnarray}
\label{EqVIIIN}
\Delta_A &=&  \frac{1}{(250 -A)} \int\limits_A^{250} d x \left| \frac{E_B^I (x) - E_B^{th}(x)}{x} \right| \,,
\end{eqnarray}
where the binding energy of set I  is $E_B^I (A)$ corresponds  to Eq. (\ref{EqIIIK})  and the  one of Eq. (\ref{EqVIInew}) with the derived  coefficients, i.e.   $E_B^{th 1}(A) = a_V (-\,A + 1.1204\,A^\frac{2}{3} +
1.1204\cdot 1.494\,A^\frac{1}{3})$.  From Table 1 (the third row from above)  one can see that the binding energy $E_B^{th 1}(A)$ with the  derived coefficients $a_S$ and $a_C$
and a single free coefficient  $a_V=16.7$ MeV   provides  a reasonable description for large nuclei with $A\ge 100$ only,  while it fails completely for $A < 50$. One more possibility is to fix the ratio $\frac{a_C}{a_S}$ (\ref{EqIX}) and vary independently the bulk and surface terms, i.e. to
employ the following parameterization 
 \begin{eqnarray}
\label{EqVIInew2}
E_B^{th 2} &\simeq& -a_V A  + a_S \left[A^\frac{2}{3} +  1.494\, A^\frac{1}{3}\right] \,.
\end{eqnarray}
For the parameters $a_V \simeq 15.8$ MeV and $a_S \simeq 14.564$ MeV with the fixed ratio $\frac{a_C}{a_S} = 1.494$ one can essentially  improve
the original fit quality as it is seen from the fourth row from above in Table 1 for  $A\ge 25$, while for $A< 25$ it still looks unsatisfactory. 
Of course, one can just ignore the derived curvature term and get a perfect description of the set I, but then one will face the problem with the Gibbs-Thomson approach to the surface tension
of spherical nuclei. 
In order to understand the source of  problem, we  have to use  the results  of refined theoretical analysis of Ref.  \cite{Kolom11}.

\begin{table*}[t!]
\begin{center}
\begin{tabular}{|l|c|c|c|c|}
                                                                                                      \hline
                 Deviation                    & $a_V$  (MeV) & $\Delta_{50}$ (MeV)   & $\Delta_{25}$(MeV)  &  $\Delta_{10}$(MeV)       \\ \hline
set I vs Eqs. (\ref{EqVIInew})-(\ref{EqIX}) & 15.60 & 1.055 &1.208 & 1.39  \\ \hline
set I vs Eqs. (\ref{EqVIInew})-(\ref{EqIX})   &           16.70 & 0.269 & 0.446 & 0.703  \\ \hline
set I vs Eq. (\ref{EqVIInew2}) \& $a_S\simeq 14.56 \,{\rm MeV}$   &           15.80 & 0.150 & 0.239 & 0.354  \\ \hline
set IV vs Eqs. (\ref{EqVIInew}), (\ref{EqXX}),  $q_3=0$  &           15.55 & 0.028 & 0.050 & 0.161  \\ \hline
set IV vs Eqs. (\ref{EqVIInew}), (\ref{EqXX}) \&  Gauss, $q_3=1.5$  &           15.50 & 0.044 & 0.073 & 0.244  \\ \hline
set I vs Eqs. (\ref{EqVIInew}),  (\ref{EqXXIIIN}), $q_3\simeq 20.5$   &           15.70 & 0.009 & 0.011 & 0.067  \\ \hline
set I vs Eqs. (\ref{EqVIInew}), (\ref{EqXXIIIN}) \&  Gauss, $q_3\simeq31.25$  &           15.70 & 0.020 & 0.026 & 0.152  \\ \hline
set IV vs Eqs. (\ref{EqVIInew}),  (\ref{EqXXIIIN2}), $q_3\simeq-3.86$   &           15.35 & 0.005 & 0.007 & 0.009  \\ \hline
set IV vs Eqs. (\ref{EqVIInew}), (\ref{EqXXIIIN2}) \&  Gauss, $q_3\simeq-4.6$  &           15.26 & 0.012 & 0.012 & 0.019  \\ \hline

\end{tabular}
\label{table1}
\end{center}
\caption{The mean deviation $\Delta_A$ per unit interval of nuclear masses between the parameterizations of binding energy obtained within  the model of  hexagonal  clusters and the  sets I and IV found in Ref. \cite{Mor:2012}. }
\end{table*}

\section{Theoretical predictions for the curvature term}

In Ref.  \cite{Kolom11} on the basis of the Gibbs-Thomson approach 
 there was developed   a refined theoretical model which accounts for many subtleties of  finite nuclei and, as a result, it successfully  determines the relation of surface and curvature terms. It employes  several parameterizations  of Skyrm interaction \cite{Skyrm1,Skyrm2} given in Ref.
 \cite{Kolom08}.  It also estimates  the Tolman length $\xi$
 via the surface tension coefficient  ($R_A\simeq r_0 A^\frac{1}{3}$ \cite{Dudek})
 \begin{eqnarray} \label{EqX}
\sigma (R_A)= \sigma_\infty \left(1-\frac{2\xi}{R_A}\right)\,, 
\end{eqnarray}
which  is negative $\xi \simeq -0.36$ fm and it weakly depends on the Skyrm model parameterization  \cite{Kolom11}.  
Eq. (\ref{EqX}) relates the surface tension $\sigma (R_A)$ of a nucleus of a radius $R_A$ with the one of infinite nucleus $\sigma_\infty$,
which has no electrical charge.  In what follows  we employ the relation $R_A\simeq r_0 A^\frac{1}{3}$  where the average value of the parameter $r_0 =1.2$ fm is taken from \cite{Dudek}. 
 
 In order to get the relation to the  coefficients $a_S$ and $a_C$ discussed above,  we have to find the free energy change from the generalized   Laplace pressure $P =  \frac{2 \sigma (R_A)}{R_A} + \frac{\partial \sigma (R_A)}{\partial R_A}$  (see Ref. \cite{Kolom11} for details):
\begin{eqnarray}
\label{EqXIN}
\Delta F (A) =  \int\limits dV  P(R) &=&  \int\limits_0^{R_A} dr \, 8 \pi r \, \sigma_\infty \left(1-\frac{\xi}{r}\right) =  4 \pi \sigma _\infty \left(R^2_A- 2\xi R_A \right)  \,.
\end{eqnarray}
Using the relation $R_A\simeq r_0 A^\frac{1}{3}$, from Eq. (\ref{EqXIN}) we find the $a_S$ and $a_C$ coefficients as 
\begin{eqnarray}
\label{EqXIIN}
a_S &=&  4 \pi \sigma _\infty \frac{R^2_A }{A^\frac{2}{3}} =   4 \pi \sigma _\infty \cdot r_0^2 \simeq 16.65 
\, {\rm MeV}\,,\\
\label{EqXIIIN}
a_C &=& - 8 \pi \sigma _\infty \frac{R_A}{A^\frac{1}{3}}\xi  = - 8\pi \sigma _\infty \,  \xi \cdot r_0   \simeq   10
 \, {\rm MeV}\,, 
\end{eqnarray}
where in the last step of evaluation we substituted  $\sigma _\infty = 0.92$  MeV$\cdot$fm$^{-2}$ found in Ref. \cite{Kolom11} for SkM  Skyrm interaction  \cite{Skyrm1,Skyrm2}  parameterized according to   Ref.  \cite{Kolom08}. This model was chosen for numerical comparison, since it is the main model of Ref. \cite{Kolom11} and since it provides the surface energy $a_S$ which is close to the one of set IV, but the other Skyrm models considered in \cite{Kolom11} give similar results. 

From Eqs. (\ref{EqXIIN}) and  (\ref{EqXIIIN}) one can find the curvature term as
\begin{eqnarray} \label{EqXIVN}
a_C \simeq 0.6 \,  a_S \,,
\end{eqnarray}
which is 40\% of  the ratio $\frac{a_C}{a_S}$ obtained within the hexagonal model Eq. (\ref{EqIX}).  
If one takes the results for SkM interaction found in a comprehensive review \cite{Brack} (see Table 8  therein), 
then one gets $\frac{a_C}{a_S} \simeq \frac{12.19}{16.6}\simeq 0.734$ for this ratio, which is about 50\%
of the value derived above in a simple way.  We believe that taking into account  the fact that the hexagonal
model does not contain any information about the complexity of nuclear interaction its findings are remarkable. 
It is, however,  appropriate to stress here that,
 to our best knowledge,  all  theoretical  estimates  of  the coefficient $a_C$  are essentially larger than the best fit with the curvature term obtained in Ref. \cite{Mor:2012} 
for the set IV,
i.e.  $\frac{a_C}{a_S}\biggl|_{sIV} \simeq \frac{3.6}{15.26} \simeq 0.24$.  

In order to  demonstrate the depth of  this problem let us  quantify the  deviation between theoretical predictions and the set IV fit of Ref.  \cite{Mor:2012} using  the Tolman length $\xi$.
The latter can be   expressed in terms of the discussed ratio  $\frac{a_C}{a_S}$.
Indeed, from Eqs.  (\ref{EqXIIN}) and  (\ref{EqXIIIN})  we  find  the Tolman length as:
\begin{eqnarray}
\label{EqXVI}
\xi &=& - \frac{a_C}{2\, a_S}   \cdot r_0 \,. 
\end{eqnarray}
The  sets I and IV  of Ref.  \cite{Mor:2012} obtained from the fit  of data  give us, respectively,  $\xi_{set I} =  0$  fm  and  $\xi_{set IV} =  - \frac{3.6}{30.5} \cdot r_0  \,\simeq - 0.14$ fm.
On the other hand, 
  similarly to essentially larger value  $\xi_{GT} = - 0.36$ fm found in Ref.  \cite{Kolom11},  the present model  gives  us $\xi_{hex} = - \frac{1.4938}{2} \cdot r_0   \,\simeq - 0.9$ fm. 
 Although the value $\xi_{set IV} \simeq - 0.14$ fm is found from the fit of nuclear binding energies, it looks too small to be the true Tolman length of large nuclei.
 In fact, the latter value means that  on the surface of nucleus its density decreases too abruptly (for details, see a discussion and figures in Ref.  \cite{Kolom11}), while the larger  theoretical values  look  more adequate from the physical point of view. Thus, we face a severe theoretical problem: on the one hand, a very successful  fit of the data corresponds to an unphysically small value of  Tolman length for large nuclei, but on the
 other hand,  the larger values of curvature term $a_C$ found in  theoretical models do not allow one to successfully describe the experimental data.
 
Therefore, the first main question addressed in  the present work can be reformulated  as follows: is  it possible to  modify the framework outlined above and the one developed in Ref.  \cite{Kolom11} (and in similar models) in order to achieve an agreement between theoretical models and the analysis of experimental data (set IV) in such a way that the models would  simultaneously provide a good description of the data and would at the same time  correspond to a physically adequate value of Tolman length?  
  
The  physically motivated and sufficiently general way to improve the treatment of light nuclei   is   to introduce the corrections to the binding  energy per pair of interacting nucleons  (ansatz I, hereafter)  which depend on the powers of $A^\frac{1}{3}$ 
\begin{eqnarray}
\label{EqXX}
\varepsilon (A) &=& \varepsilon_0 \left[ 1 + \frac{q_1}{A^\frac{1}{3} } + \frac{q_2}{A^\frac{2}{3}} + \frac{q_3}{A^1} + ... \right]  \,,
\end{eqnarray}
where the parameters $q_1$,  $q_2$ and $q_3$ are the constants which should be determined from the best description
of the  experimental data, whereas  the uncorrected binding energy of the nucleon pair is $\varepsilon_0$.
   Note that the ansatz I is just  the leptodermous expansion. It is well known that these corrections naturally
appear for finite nuclei from such an  expansion \cite{Brack}, but compared to the approach used in Ref. \cite{Brack} the advantage  of the ansatz I is that it modifies all  coefficients $a_V$,  $a_S$, $a_C$ and $a_G$ simultaneously, thus, keeping the minimal number of fitting parameters.

\begin{figure}[t]
\centerline{\includegraphics[height=99mm,width=99mm]{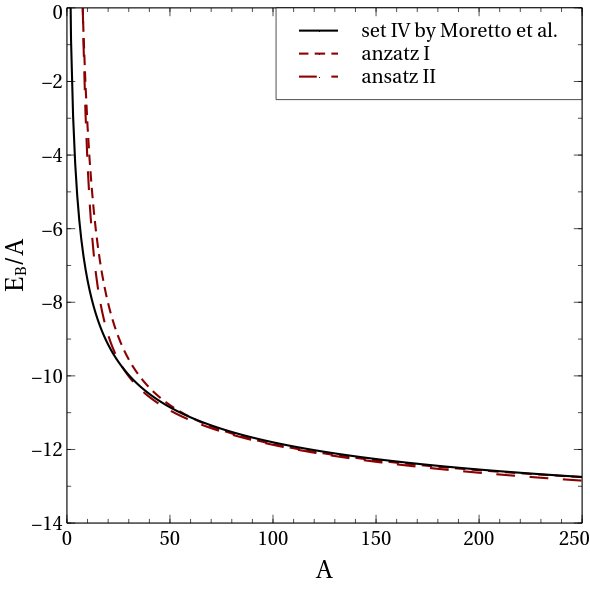}}
\caption{Comparison of the binding energy per nucleon of set IV  from Ref. \cite{Mor:2012} (solid black curve) with the hexagonal model
with ansatz I  (short dashed curve) and ansatz II (long dashed curve). For more details see the text.}
\label{fig3}
\end{figure}

Apparently,  such an assumption   allows us  to compensate the most part of curvature term by the proper choice of 
parameters $q_{k}$. Assuming that  $q_{k\ge3} =0$ and substituting  Eq. (\ref{EqXX}) into expression for the derived binding energy (\ref{EqVIInew}),   one can express the coefficients $q_1$ and $q_2$ via the coefficients $a_{\{K\}}$:
\begin{eqnarray}
\label{EqXXI}
q_1 =  \frac{5}{2}\left[\frac{3}{10} \right]^\frac{2}{3} - \frac{ a_S}{a_V} \simeq 0.1204 \,, ~ q_2 =  \frac{5}{2} q_1 \left[\frac{3}{10} \right]^\frac{2}{3} +
\frac{5}{2}\left[\frac{3}{10} \right]^\frac{1}{3}  - \frac{ a_C}{a_V} \simeq 1.5725 \,.
\end{eqnarray}
However, it is more instructive to consider the parameters $q_1$ and $q_2$ as the fitting parameters to refine the derived expression for binding energy (\ref{EqVIInew}).   As one can see from  Fig. \ref{fig3}, the parameters   $q_1$ and $q_2$ given in the second  row of Table 2 provide essentially better description of the set IV. A quantitative analysis shows (see Table 1) that the  deviation $\Delta$
calculated for the
 ansatz I is about one order of  magnitude  smaller, than without it.  Visually, from Fig. \ref{fig3} one can see that Eq. (\ref{EqXX}) 
allows us to perfectly reproduce  the set IV for $A\ge 40$, and  reasonably well for $A\in [20, 40]$, while for  $A\le 20$
this ansatz fails.  
Usually, for ordinary nuclei such corrections are playing an auxiliary role, but for studying the nuclear matter 
properties they are important. To show this let us  estimate the  values for the apparent coefficients 
\begin{eqnarray}
\label{EqXXIb}
a_S^{appI} &=& a_S - a_Vq_1 \simeq 1.444 a_V  \simeq 22.46 ~{\rm MeV}\, ,\\
\label{EqXXIc}
a_C^{appI} &=&  a_C - a_Vq_2 + a_S q_1 \simeq -2.758 a_V  \simeq -42.88 ~{\rm MeV}\, ,\\
a_G^{appI} &=&  a_G - a_Vq_3 + a_S q_2 +a_Cq_1 \simeq 4.021 a_V  \simeq 62.52 ~{\rm MeV}\, ,
\label{EqXXId}
\end{eqnarray}
which are obtained after reordering the terms of binding energy, if one accounts for 
  the corrections  given in  Eq. (\ref{EqXXI}).  These numbers are obtained for the coefficients given in the second row 
of Table 2 assuming $a_G=0$. From this example one can see that the  corrections given in  Eq. (\ref{EqXXI}) may essentially modify the binding energy of large nuclei. It is an interesting question whether the apparent coefficients   (\ref{EqXXIb}), (\ref{EqXXIc}) and (\ref{EqXXId}) used within the SMM 
will essentially modify its results or not, since the larger value of  $a_S^{app}$ coefficient maybe compensated by negative value of 
the $a_C^{app}$ coefficient.

One can also reproduce the set IV with the  same quality, if the Gauss curvature term, $-\frac{27}{2} \varepsilon = \frac{27}{24} a_V$ or the last term in Eq. (\ref{EqV})   is taken into account.  The corresponding parameters are given in the third row of Table 2. 

\begin{table*}[t!]
\begin{center}
\begin{tabular}{|l|c|c|c|c|}
                                                                                                      \hline
                Ansatz                   &~  $a_V$  (MeV) ~&~ $q_1$  ~  &~ $q_2$ ~ & ~ $q_3$ ~      \\ \hline
I:  Eqs. (\ref{EqVIInew}),  (\ref{EqXX}) & 15.55 & -0.324 & 4.075 & 0  \\ \hline
I:  Eq. (\ref{EqVIInew}) \&  Gauss, Eq. (\ref{EqXX})   &           15.50 & -0.494 & 5.05 & 1.5  \\ \hline
II: Eqs.(\ref{EqVIInew}), (\ref{EqXXIIIN})    &           15.70 & 0. & 1.43 & 20.5  \\ \hline
II: Eq. (\ref{EqVIInew}) \&  Gauss,  Eq. (\ref{EqXXIIIN})   &           15.70 & 0. & 1.35 &  31.25  \\ \hline
III: Eqs.(\ref{EqVIInew}), (\ref{EqXXIIIN2})    &           15.35 & 0. & 2.3 & -3.86  \\ \hline
III: Eq. (\ref{EqVIInew}) \&  Gauss,  Eq. (\ref{EqXXIIIN2})   &           15.26 & 0. & 2.65 & -4.604  \\ \hline
\end{tabular}
\label{table2}
\end{center}
\caption{Different sets of  parameters used to modify the hexagonal model results in order to better reproduce the sets I and IV.
The details can be found in Table 1. 
The first column refers to the corresponding parameterization.  The second, the fourth and the sixth rows do not account for the Gauss curvature term, while the third, the fifth and the seventh ones account for it.}
\end{table*}
{
The reason of why a simple leptodermous-like correction of Eq. (\ref{EqXX}) fails to reproduce the binding energy for $A< 20$ is apparent. At small $A$ values the correction provided by Eq. (\ref{EqXX}) is not small and, hence, it dramatically  changes the $A$-dependence of expression  $(\ref{EqVIInew})$. In order to avoid such a problem for the nuclei with  masses in the range $A \in [10; 50]$ we employ a less sophisticated form of correction, the ansatz II afterwards, 
\begin{eqnarray}
\label{EqXXIIIN}
\varepsilon (A) &=&  \varepsilon_0 \left[ 1 + \frac{q_2}{A^\frac{2}{3} } + \frac{q_3}{A^\frac{4}{3}} \right]  \,, 
\end{eqnarray}
which has two parameters only, but  a higher $A$-power of  the term next to the $A^{-\frac{2}{3}}$ correction.
Its advantage is that by construction such an  ansatz does not affect the surface coefficient $a_S$, but modifies the higher order terms.
We also analyzed the two parametric version of  Eq.  (\ref{EqXX}) with $q_1 =0$, when the parameters   $q_2$ and $q_3$ were used to fit 
the set I,  and found that the ansatz II provides an essentially better 
description of the sets I and IV. 
As one can see from Table 1, compared to the ansatz I,  this ansatz with one less parameter provides a  better  description of the set I  for $A\ge 10$ both without the Gauss curvature term and with it.  The values of corresponding parameters are given in Table 2. For the case $a_G=0$ the ansatz II generates the following apparent coefficients for nuclear matter
\begin{eqnarray}
\label{EqXXIIIb}
a_S^{appII} &=& 1.1204\, a_V  \simeq   17.44 ~{\rm MeV}\, , \\
\label{EqXXIIIc}
a_C^{appII} &=& a_C - q_2\, a_V \simeq  0.24\, a_V  \simeq 3.82~{\rm MeV}\, , \\
\label{EqXXIIId}
a_G^{appII} &=& q_2\, a_S^{appII} \simeq   1.602\, a_V  \simeq 25.15 ~{\rm MeV}\, .
\end{eqnarray}
Comparing these coefficients with the ones found for the ansatz I, one can conclude that (i) their values strongly depend on the  quality of light nuclei description; (ii) the ansatz II coefficients $a_V$ and $a_S^{appII}=1.1204\, a_V $   almost coincide with the ones of the set I and  with the usual SMM values \cite{SMM}, while  the curvature coefficient $a_C$ almost matches  the corresponding coefficient  of the set IV; (iii) for the both ansatze the Gauss curvature coefficient is sizably larger   than  the bulk one.  
{In Fig. \ref{fig3} we on purpose compare the ansatz II with the set IV, although its parameters were obtained from the fit of set I. This is done in order to demonstrate that for a high quality fit a tiny difference between the sets I and IV observed at $A > 50$ matters. }
 
Including in the treatment the Gaussian curvature term $-\frac{27}{24} a_V$,  we obtain no change for the bulk and surface coefficients,  only a slight numerical shift  for the coefficient $q_2$ and an essential  increase of the coefficient $q_3$. 
Consequently,  the coefficient
$a_C^{appII} \simeq 5.07$ MeV is increased only  by 25\%, while  the coefficient  $a_G^{appII}   \simeq 41.41$ MeV gained the contribution from the  Gaussian term $\frac{27}{24}\,a_V$ of hexagonal cluster model.  Thus,  ansatz II not only provides the best description of the nuclear binding energy with three parameters only, but, compared to the ansatz I,  it  also demonstrates a stability of the  apparent values of 
all coefficients.  Therefore, we consider the ansatz II as the most successful one. 
}

{
This ansatz  perfectly matches the sets I and IV for $A> 20$, but for $A= 10-14$ nucleons the typical deviation from these sets is about 3-6 MeV per nucleon. Therefore now we concentrate on improving the description of light nuclei with  the minimal number of parameters. 
To  complete this task and to demonstrate that a better  quality of light nuclei  (with 10-20 nucleons) description  may 
essentially affect the apparent values of the curvature and Gauss terms, we consider  the correction with two parameters, the ansatz III, 
\begin{eqnarray}
\label{EqXXIIIN2}
\varepsilon (A) &=& \varepsilon_0 \left[ 1+ \frac{q_2}{A^\frac{2}{3}+ q_3}\right] \underbrace{\longrightarrow}_{A \gg 10} \varepsilon_0 \left[1+  \frac{q_2}{A^\frac{2}{3}} - \frac{q_2 \, q_3}{A^\frac{4}{3}} + ... \right]  \,,
\end{eqnarray}
where in the limit $A\gg 10$ we expanded the denominator to get an asymptotic form of the ansatz III.
As one can see from Table 1 this ansatz provides the best description of the set IV for $A\ge 10$ both without the Gauss curvature term and with it although it has a pole at $A < 10$. The values of corresponding parameters are given in Table 2. For the case $a_G=0$ this gives  the  apparent values of  the nuclear matter coefficients 
\begin{eqnarray}
\label{EqXXIIIb2}
a_S^{appIII} &=& 1.1204\, a_V  \simeq   17.2 ~{\rm MeV}\, , \\
\label{EqXXIIIc2}
a_C^{appIII} &=& a_C - q_2\, a_V \simeq - 0.627\, a_V  \simeq -9.62~{\rm MeV}\, , \\
\label{EqXXIIId2}
a_G^{appIII} &=& q_2\, a_S^{appII} \simeq   2.577\, a_V  \simeq 39.56 ~{\rm MeV}\, .
\end{eqnarray}
Although this ansatz  perfectly matches the sets I and IV for $A\ge 10$, one can see that the found value of the apparent curvature coefficient 
does not reproduce either the set I  value $a_C=0$ or  set IV result  $a_C=3.6$ MeV.  
If the Gaussian curvature term $-\frac{27}{24} a_V$ is included into the treatment,  one finds 
that  the  coefficients
$a_C^{appIII} \simeq -14.9$ MeV and $a_G^{appIII}   \simeq 62.93$ MeV are increased in about  one and half  times, i.e. in contrast to the ansatz II  both of these coefficients are  modified essentially. 

 Of course,  we  analyzed the other forms  of corrections and surprisingly found that the ones, which have simple pole  provide a better description of the sets I and IV
with a smaller number of parameters. In particular, despite an inadequate  behavior at $A< 10$ the ansatz III  provides the best description of set I with the minimal number of parameters. We believe that for small nuclei with less than 20 nucleons the present model cannot be  applied and the point  that the derived parameterization of binding energy has to be supplemented by the corrections with the pole is a reflection of the fact that
the binding energy of  small nuclei should be corrected  differently than it is done for the larger ones. 
}

A similar way to improve the coefficient  $a_C$  given by Eq. (\ref{EqXIIIN})  cannot, however, 
be used, since it is already  obtained from the leptodermous expansion of the Skyrm interaction \cite{Kolom11}. Our educated guess is that one possible solution of this problem is related to the fact that the radius of nucleus of $A$ nucleons can be modified as 
\begin{eqnarray}
\label{EqXXIVN}
R(A) &=& r_0 A^\frac{1}{3} \left[ 1+  q_1 A^{-\frac{1}{3}} \right] \,.
\end{eqnarray}
In this case the free energy (\ref{EqXIN})  generated by the Laplace pressure can be written as 
\begin{eqnarray}
\label{EqXXVN}
F (A) =    4 \pi \sigma _\infty \left[R^2(A)- 2\xi R(A) \right] \simeq 4 \pi \sigma _\infty \left[ r_0^2 A^\frac{2}{3} + 2( r_0 q_1-  \xi) r_0 A^\frac{1}{3} + ...  \right]\,,
\end{eqnarray}
where on the right hand side we neglected the terms proportional to $A^0$. Apparently, such an assumption allows one to reduce   the  coefficient $a_C$ to its value found in Ref. \cite{Mor:2012} without modifying the true Tolman length, if   one chooses $q_1 = \frac{\xi_{GT}-\xi_{setIV}}{r_0} \simeq -0.183$. 

The suggested approach allows us to easily  elucidate the temperature dependence of the coefficients $a_S$, $a_C$ and $a_G$, if the corresponding 
dependence of the bulk term $a_V (T)$ is known. Taking the usual SMM parameterization of the bulk term as $- a_V (T)  = - a_V (0) + \frac{T^2}{\epsilon_F} $ (where $\epsilon_F \simeq 16$ MeV \cite{SMM}) which accounts for the Fermi motion of nucleons at non-vanishing  temperature $T$, one can find
\begin{eqnarray}
\label{EqXXVIN}
a_B(T) = a_V(T) \frac{a_B(0)}{a_V(0) }\,, ~{\rm with}~ B \in \{S, C, G\} \,,
\end{eqnarray}
since by construction all these coefficients can be expressed in terms of  temperature dependent binding energy per pair of nucleons 
$\varepsilon = -\frac{a_V(T)}{12}$.  At the temperature 
\begin{eqnarray}
\label{EqXXVII}
T_c  = \left[ a_V(0) \epsilon_F \right]^\frac{1}{2} \in [15.6; 16.4] ~{\rm MeV} \,,
\end{eqnarray}
the surface tension coefficient $a_S(T)$ together with the bulk and curvature ones vanish  and, hence,  at this temperature and above it  the stable clusters cannot exist. The obtained range of $T_c$ in Eq. (\ref{EqXXVII}) is found by taking the values of the bulk coefficient $a_V(0)$ given in the second column of Table 1. It is remarkable that such a range of $T_c$ values  is consistent with 
the values of the critical endpoint (CEP) temperature found for the nuclear matter  from the analysis of experimental data \cite{Tcep2006}.
Also this range of $T_c$ values  fits very well  into  the estimates of CEP temperature obtained 
within the mean-field equation of state  with the  realistic  hard-core repulsion between the nucleons which allows one to go beyond the popular Van der Waals approximation \cite{IST17a}. Moreover, as one can see from Fig. 3 of Ref.  \cite{IST17a} such a range of CEP temperatures is consistent with the range of the nuclear matter incompressibility constant $K_0 \in [270; 315]$ MeV determined  recently in 
 \cite{Stone:2014,Wang:2018}. 

Note that  an explicit  temperature dependence of the surface tension coefficient 
\begin{eqnarray}
\label{EqXXVIII}
a_S (T) = 1.1204 \left[a_V(0) - \frac{T^2}{\epsilon_F} \right] = a_S(0) \left[1 - \frac{T^2}{T_c^2} \right]   \,,
\end{eqnarray}
is, on the one hand, absolutely identical to the $T$-dependence of surface coefficient   deduced  in \cite{Wagner:2001} from the analysis of nuclear multifragmentation data. 
Clearly, the temperature dependence of the curvature coefficients is the same as  in Eq. (\ref{EqXXVIII}). Due to the fact that  the  corrections given by Eq.  (\ref{EqXX})  or Eq. (\ref{EqXXIIIN})  have the same $T$-dependence, then such a  $T$-dependence should be also valid for the corresponding  apparent coefficients  defined  by Eqs. (\ref{EqXXIb})-(\ref{EqXXId}) or by Eqs. (\ref{EqXXIIIb})-(\ref{EqXXIIId}). 
Although the $T$-dependence of $a_S (T) $ coefficient 
(\ref{EqXXVIII}) was criticized in \cite{Tcep2006}  as poorly consistent with the multifragmentation data, we should stress that the analysis of Eq.  (\ref{EqXXVIII}) made in \cite{Tcep2006} does not take into account the presence of the curvature terms and, hence, it cannot 
be applied to the framework suggested here.

Besides, in the vicinity of $T_c$ the surface tension coefficient $a_S (T)$ (\ref{EqXXVIII}) linearly depends on temperature which is not only 
similar  to the Fisher droplet model \cite{Fisher}, but also to the result obtained within the exactly solvable model of surface partition 
of large physical clusters \cite{HDM1, HDM2}.
 Therefore, the found values of $T_c$ may be considered as a realistic estimate for
the CEP temperature of nuclear matter.

Of course, it is possible that the $T_c$ values shown above will get slightly larger, if one takes into account the surface tension 
induced by the repulsive and attractive   interaction of  large nuclear matter clusters  with the thermal medium \cite{IST17a,IST:2014}, but to estimate such effects one needs more elaborate model than the present one.


\section{Conclusions and perspectives}

Based on the saturation property of nuclear interaction we developed here a hexagonal model of large nuclei in which all the coefficients 
of leptodermous expansion of nuclear binding energy  are expressed  in term of  the  binding energy of  a pair of nearest neighboring nucleons.  
It is remarkable that at zero temperature such a model  reproduces the asymptotic ratio   surface to bulk binding energy coefficients known from the traditional Bethe-Weizs$\ddot{\rm a}$cker formula \cite{BW}, and that  with the deviation of $40-50$\% it also reproduces the ratio of the curvature to surface tension leptodermous coefficients obtained within sophisticated parametrization of Skyrm interaction between nucleons \cite{Brack,Kolom11}. 
In addition, the suggested approach allows one to derive the Gauss curvature term.  In our opinion, this is very good result for the geometrical model which
does not contain any information about the complexity of nuclear interaction. 

The two major advantages of the hexagonal model are that it is very simple and that  it does not rely on the value of  the nuclear density of ground state and on  the relation between the radius of nucleus $R(A)$ and the number of nucleons $A$ in it. Due to these  advantages we were able to simultaneously determine the corrections to the surface and curvature coefficients and estimate their  apparent values which should be used to evaluate the properties of symmetric nuclear matter. 
{Surprisingly,  the leptodermous-like corrections (ansatz I) do not provide the best result even for a larger number of parameters.  
Excluding the ansatz III which has a pole at  number of nucleons below 10,
the best correspondence to the data is provided by the ansatz II 
whose bulk and surface coefficients are very close to the usual SMM values.  However,  in addition this ansatz   generates the both curvature terms which should be included into the SMM and studied in details.}
Furthermore, the advantages  of the suggested model allowed us to express the temperature dependence of 
all these coefficients in terms of the temperature dependent bulk one.  The found range of critical temperature $T_c$ obtained for a single  large nucleus in a vacuum is in a very good agreement  with theoretical and experimental values of this quantity.  Of course, in addition to the values of
proper surface and curvature  coefficient discussed here  one has to account for their modification in the thermal medium in a spirit of approach suggested in  \cite{IST14}. 

It is clear that for a more reliable determination of the apparent values of the coefficients $a_V$, $a_S$, $a_C$ and $a_G$ from the experimental binding energies one has to take into account all the terms in Bethe-Weizs$\ddot{\rm a}$cker formula, but in this case the hexagonal model should be also improved by considering a more realistic interaction between nucleons.  Furthermore, we believe that such an approach will be 
interesting to estimate the properties of  molecular clusters in real gases, but such an analysis is out of the scope of present work. \\

{\bf Acknowledgements.} The authors are thankful to O. M. Gorbachenko, 
B. E. Grinyuk, V. Yu. Denisov, A. P. Kobushkin, I. N. Mishustin and  L. M. Satarov for fruitful discussions and valuable comments. 
V.V.S. thanks the Funda\c c\~ao para a Ci\^encia e Tecnologia (FCT), Portugal, for the
financial support through the grant No. UID/FIS/04564/2019. 
The work of  V.V.S., K.A.B., A.I.I.  was supported in part  by the Program of Fundamental
Research of the Department of Physics and Astronomy of the National Academy of Sciences of Ukraine (project No. 0117U000240).  The work of A.I.I. was done within the project SA083P17 of Universidad de Salamanca launched by the Regional Government of Castilla y Leon and the European Regional Development Fund. 
K.A.B. is grateful to the COST Action CA15213 ``THOR'' for supporting his networking.
Also K.A.B. is grateful for a warm hospitality to the colleagues from the university of Oslo,   where this work was completed. 
 


\begin{thebibliography}{99}


\bibitem{SMM}
%
J. P. Bondorf, A. S. Botvina, A. S. Iljinov, I. N. Mishustin, and K. S. Sneppen, Phys. Rep. 257, 131 (1995) and references therein.

\bibitem{Fisher}
 %
M. E. Fisher,  Physics {\bf 3}, .255 (1967). 

\bibitem{SMM1}
 %
S. Das Gupta and A.Z. Mekjian, Phys. Rev. C 57, 1361 (1998).


\bibitem{SMM2}
K. A. Bugaev, M. I. Gorenstein, I. N. Mishustin  and  W. Greiner,
Phys. Rev. C {\bf 62}, 044320  (2000); and
%
 Phys. Lett. B {\bf  498}, 144 (2001).

\bibitem{SMM3}
 %
P. T. Reuter and K. A. Bugaev,
 Phys. Lett. B {\bf  517},  233 (2001).

\bibitem{IST14}
 %
 V. V. Sagun, A. I. Ivanytskyi, K. A. Bugaev and I. N. Mishustin,
Nucl. Phys. A {\bf 924},  24 (2014).

\bibitem{IST18b}
 %
V. V. Sagun {\it et al.,}
 Eur. Phys. J. A {\bf 54},  100 (2018).
 
 \bibitem{IST17a}
 %
  A.~I.~Ivanytskyi, K.~A.~Bugaev, V.~V.~Sagun, L.~V.~Bravina and E.~E.~Zabrodin,
 Phys. Rev.  C {\bf 97},  064905 (2018). 

\bibitem{Axioms1}
%
L. van Hove. 
Physica {\bf 15}, 951 (1949).

\bibitem{Axioms2}
%
L. van Hove. 
Physica {\bf  16}, 137 (1950).

\bibitem{BW}
%
C. F. von Weizs$\ddot{\rm a}$cker,  Z. Phys. {\bf 96}, 431 (1935).

\bibitem{BW2}
%
W. D. Myers and W. J. Swiatecki, Nucl. Phys. {\bf 81}, 1 (1966).

\bibitem{BW3}
%
W. D. Myers and W. J. Swiatecki,  Nucl. Phys. A {\bf 601}, 141 (1996).

\bibitem{Brack}
M. Brack, C. Guet and H.-B. H/\hspace*{-2.2mm}okansson, Phys. Rep. {\bf 123}, 276 (1984)
and references therein.

\bibitem{Dudek}
K. Pomorski and J. Dudek Phys. Rev. C {\bf 67}, 044316 and references therein.

 \bibitem{Kolom08}
V. M. Kolomietz and A. I. Sanzhur, Eur. Phys. J. {\bf 38}, 345 (2008).

\bibitem{Mor:2012} 
L. G. Moretto, P. T. Lake and L. Phair,
Phys. Rev. C {\bf 86}, 021303(R) (2012) and references therein.



\bibitem{Kolom11}
 V. M. Kolomietz, S. V. Lukyanov and A. I. Sanzhur,
 Phys. Rev. C {\bf 86},   024304 (2012) and references therein.

\bibitem{Curvat53}
D. L. Hill and J. A. Wheeler, Phys. Rev. {\bf 89}, 1102 (1953). 

\bibitem{NewFisher1}
A. Dillmann and G.E. Meier, J. Chem. Phys. {\bf  94}, 3872  (1991).


\bibitem{NewFisher2}
A. Laaksonen, I.J. Ford and M. Kulmala, Phys. Rev. E {\bf 49}, 5517  (1994) and references therein.

\bibitem{Cells}
J. G. Kirkwood  and F. P. Buff,  J. Chem. Phys. {\bf 17},
338 (1949).

\bibitem{FiniteT}
D. G. Ravenhall, C. J. Pethick, and J. M. Lattimer, Nucl. Phys. A {\bf  407}, (1983) 571.

\bibitem{Mack:1962} 
A. L. Mackay, 
  Acta\ Cryst. {\bf 15}, 916 (1962).

 
 \bibitem{Skyrm1}
  T. H. R. Skyrme. 
   Phil. Mag. {\bf 1}, 1043 (1956).
   
  \bibitem{Skyrm2}
  T. H. R. Skyrme. 
  Nucl. Phys. {\bf 9}, 615 (1959). 



 
 \bibitem{Tcep2006} 
  V. A. Karnaukhov, Phys. Part. Nucl. {\bf 37}, 165 (2006) and
references therein.
  
 \bibitem{Stone:2014}
 %
J. R. Stone, N. J. Stone, and S. A. Moszkowski, Phys. Rev. C {\bf 89}, 044316 (2014).  

 \bibitem{Wang:2018}
 %
 Y. Wang {\it et. al.,},  Phys. Lett. B {\bf 778},  207 (2018). 

\bibitem{Wagner:2001}
 %
J. Richert  and P. Wagner,
Phys. Rep.  {\bf 350}, 1 (2001).
 

 
  \bibitem{HDM1}
 %
K. A. Bugaev, L. Phair and J. B. Elliott,
 Phys. Rev. {\bf E 72},  047106  (2005).
 
\bibitem{HDM2}
 %
 K. A. Bugaev  and J. B. Elliott,
Ukr. J. Phys. {\bf 52}, 301   (2007).

\bibitem{IST:2014}
 %
V. V. Sagun, A. I. Ivanytskyi, K. A. Bugaev and I. N. Mishustin,
Nucl. Phys. A {\bf 924}, 24   (2014),


\end{thebibliography}
\end{document}